# A Survey of Current Trends in Distributed, Grid and Cloud Computing

GAURAV MITTAL[1], Dr. NISHTHA KESSWANI[2], KULDEEP GOSWAMI[3]
[1,2]Central University of Rajasthan, Bandra Sindri, Ajmer
[3]Government Women Engineering College, Ajmer

## Abstract

Through the 1990s to 2012 the internet changed the world of computing drastically. It started its journey with parallel computing after it advanced to distributed computing and further to grid computing. And in present scenario it creates a new world which is pronounced as a Cloud Computing [1]. These all three terms have different meanings. Cloud computing is based on backward computing schemes like cluster computing, distributed computing, grid computing and utility computing. The basic concept of cloud computing is virtualization. It provides virtual hardware and software resources to various requesting programs. This paper gives a detailed description about cluster computing, grid computing and cloud computing and gives an insight of some implementations of the same. We try to list the inspirations for the advent of all these technologies. We also account for some present scenario faults of grid computing and also discuss new cloud computing projects which are being managed by the Government of India for learning. The paper also reviews the existing work and covers (analytically), to some extent, some innovative ideas that can be implemented.

## I. Introduction

A few years ago to access the information and use resources a user required infrastructure and computer system on same location like a data storage devices. Grid computing came in a mid-1990s with a goal to provide an opportunity to the user to remotely utilize ideal computing power within other computing centres when the local computing centres are busy [4]. Then researchers developed their idea to provide not only computing power on demand but also provide data, storage devices and software's on demand. Then the term cloud computing came to deliver reliable, flexible and minimum cost of resources to third person [13]. Cloud computing came by the end of 2007. But after the advent of cloud, it removed the need of infrastructure on the same location. We can access any information from anywhere anytime. Virtualization is the main building block of Cloud Computing. It virtually provides a resource on demand which are priced. Only internet connection is necessary for accessing the services of cloud. The major feature of cloud is that we can access the same document from anywhere by any device which has the capability to access the internet. It is very beneficial for small organizations which cannot afford heavy infrastructure and storage devices. The small companies and organizations can store their information on cloud by removing the cost of purchasing the storage devices. Cloud computing systems combine the virtualization, automated provisioning and internet connectivity technologies.

## II. Cluster Computing

Cluster is a merging of parallel or distributed processing system which consists thousands of stand-alone computers which work cooperatively as a single high speed computer system. Clusters are mainly used for high availability, load balancing and for higher computing purpose.

Since some of the problems in the field of science, engineering, and business could not be solved by supercomputers, they were attempted on clusters with the goal to overcome the problems of high computation and the cost of solving them by supercomputers [4]. The components of a cluster are connected to each other through fast local area network. In cluster architecture multiple computers are connected together in a form of cluster and work as a single virtual computer which shares its workload. When user works, the request are received by the server and distributed to all the standalone computers for computing. The result is sharing workload among thousands of standalone computers for faster execution and higher computing. However cluster is characterised into three major parts High availability cluster, Load balancing cluster, and HPC cluster. The benefits of clusters are scalability, availability and high performance.

One important cluster application is Google search engine, Petroleum reservoir simulation, protein explorer, Earthquake simulation, and Image rendering.

Some important challenges in cluster computing are elasticity, middleware, scalability, programme, security and encryption, load balancing, manageability, etc. Clusters are mainly designed for databases applications. For example in windows it can be an active directory cluster, and on web programming it can be My SQL database cluster.

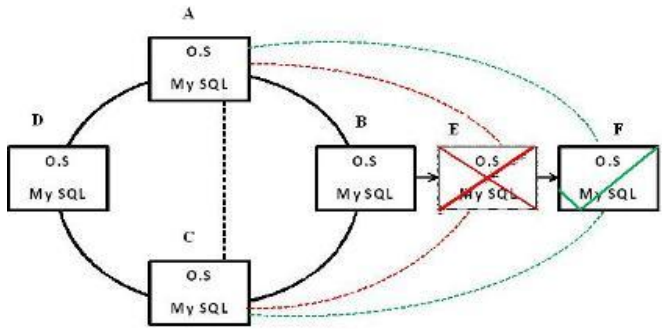

**Figure 1: Cluster computing architecture**

As we can see in the figure 1 here are four boxes A, B, C, D which are directly connected to each other. The feature of a cluster is that it does replication. So every computer in a cluster has same information as other computers have. Here all boxes have an operating system and My SQL database installed on every computer. Suppose the user coming through the internet accesses the database residing on computer D. but computer D has already enough applications so it will automatically be redirected to computer A, B or C. because of the replication. It is a main concept of cluster. But suppose computer B has failed, which is indicated in the figure by E. When a computer B fails the A will automatically break connection to B and will directly connect to C. and when the computer B will be up (in figure it is F) it will again connect to B.

## III. Grid Computing

Grid computing is simply stated as a next step of distributed computing. Grid computing was inspired by the electrical grid or power grid. Wall outlets of the electrical grid help in connecting to the resources and generate and distribute the bills of electricity. When the user connects to the power grid he does not need to know the physical location of the power plant. All a consumer bothers about is that he is getting the required amount of power for use. Same philosophy is followed by grid computing, it uses middleware layer to coordinate different IT resources across a network, which works as a virtual whole.

The aim of a computing grid is same as that of an electrical grid i.e., which provides resources to the user when they need them.

The goal of a grid computing is to create an illusion of large and powerful self-managing virtual systems out of a large collection of connected heterogeneous systems sharing various combinations of resources [3].

It uses a layer of middleware to communicate with the heterogeneous hardware and data set. Grid computing is based on internet protocols and ideas of parallel and distributed computing. It provides sharing of Computational resources, storage elements, specific applications, and equipment etc. [6]. Two main goals of grid computing first is to decrease the total elapsed computation time while leveraging existing hardware and the second is maximum utilization of unused computing and hardware resources [5]. Grid uses some communities called virtual organizations to share geographically distributed resources to achieve a common goal. In grid computing virtual organization is defined as "a group of individual or institutions who share the computing resources of a grid for a common goal" [7].The applications which are commonly used by personal computers are SETI@home.project, Screensaver lifesaver, and climateprediction.net, LHC. Some domains of a grid computing are high-energy physics, biology, Earth science, Astrophysics, Fusion, Computational chemistry etc. Some important features characteristics which are provided by computational grids are heterogeneity, scalability, adaptively or fault tolerance, and security. The major components that are necessary to form a grid are user level, middleware level, and resource level.

## IV. Present scenario grid fault

An electricity grid is a network of power lines which evaluates electricity from a generating station. India's electricity grids are divided into five regional zones, north, east, south, west and north-east to optimally utilize the unevenly distributed power resources in the country.  As we all are know that $30^{th}$ and 31th July 2012 Northern grid has failed. Due to this outage over 620 million people affected, it is nearly a 9% of the world population or half of the Indian population. It affected across 22 states in Northern, Eastern and Northeast India.

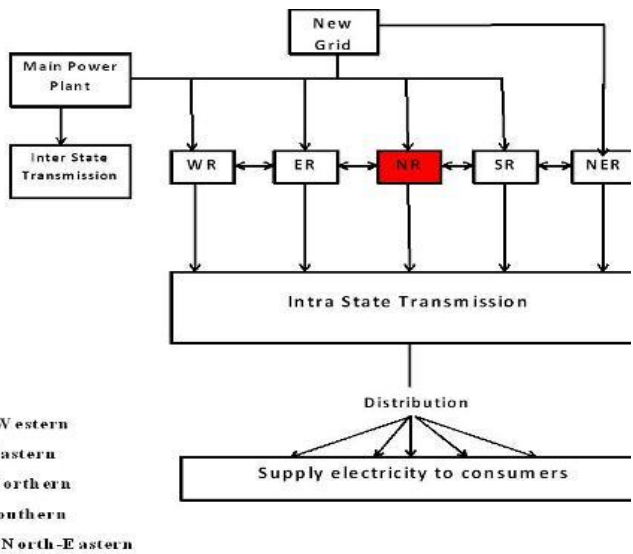

Figure: 2 Electrical grid layout

On 30th July the generation plant and all the power grids were working in a synchronized manner. All the zones WR, ER, NR, SR, and NER do load balancing. On warning issues all power grids balanced their work load. The permissible limit of which was defined by the Indian electricity grid code is 49.5 Hz to 50.2 Hz. When the warning came to the main power plant the northern zone again consumed electricity from it. Due to exhausted power all the zones were unable to balance the load and as a result the northern grid failed. And it also affected the north eastern and eastern power grids.

We are discussing the power grid because it has attained enough saturation and its analogy can be used for high performance computing and hence, disasters of this sort are also obvious. So, the advocates of Grid Computing will have to find an answer to this sort of a situation which we consider can be the future work in this area.

## IV. NKN Project

The "National knowledge network" is GOI approved project. The goal of this project is to connect all universities and institutions for higher learning and research with a 10 Gbps data communication network to facilitate knowledge sharing and collaborative research. Currently it is working with 7 super nodes and 24 core nodes which connect 628 institutions and remaining 850 institutes will be connected by March 2013. This project was initiated by the Government of India and it is a grid computing project for using ideal resources and data of institutions. It is biggest step in e-learning and many of the application areas are Agriculture, Education, Health, and E-governance.

## Cloud Computing

Cloud computing is considered as an evolution of grid computing to extract more or increase the infrastructure-based services.

The cloud is a new service which is used for delivering resources like computing and storage to customers on demand [9].

Cloud computing provides highly scalable computing resources as services on payment basis. A major application of cloud computing is that consumer only uses those services which he needs. Resources are available for accessing at 24*7 and consumer can access it from any location via internet. We don't need to worry about how servers or resources are maintained behind the scenes, we simply purchase resources according to need as on a rent/lease basis. Cloud computing has also been called utility computing or 'IT on demand'. It is a new business model which uses latest IT technologies like virtualization and multi-tenancy. Both these services are used to take advantages of economics of scale and to reduce the cost of IT resources. General example of cloud services is Google apps serviced by Google and Microsoft SharePoint [11].

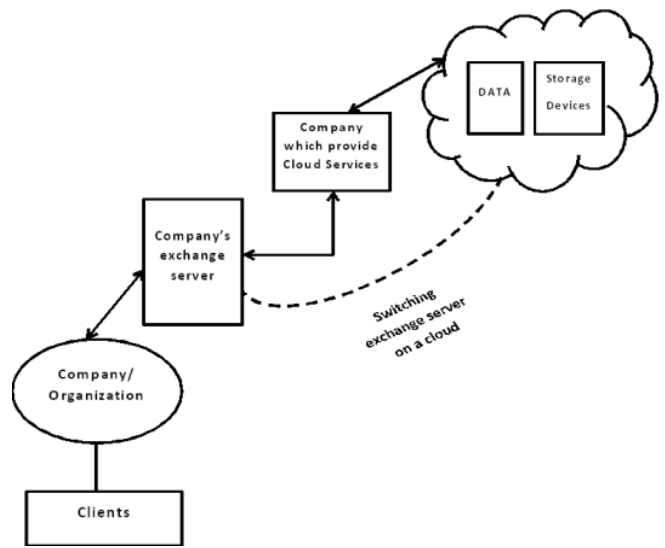

Figure: 3 cloud computing mechanism

We now present a possible application of cloud computing. Figure illustrates that here is a company which uses an exchange server for private mailing service. Suppose this organization has 200 hosts and they are using services of exchange server. If a server is down or fails then company's work will stop and it will be a great loss in terms of money and resources. For getting rid of this type of disaster we use a cloud services.

As we can see from the figure company's exchange server is switched on a cloud by a cloud service provider on pay-as-service. Now if a server fails or down company has no worry about it and the server will always run by a backup services of a cloud. This is a basic fundamental of cloud computing. Here is three types of cloud computing 1). Public cloud 2). Private cloud 3). Community cloud 4). Hybrid cloud. The above example of private cloud. If two or more organizations uses a same cloud than it is called a community cloud and a public cloud is accessed by a subscriber with an internet connection. The major features of cloud are Elasticity, Reliability, Virtualization, Quality of service, Agility and adaptability etc. The benefits of cloud computing are Cost reduction, Ease of use.

## V. Comparative Study of Cluster, Grid and Cloud Computing

In a conceptual manner cluster, grid and cloud computing contains a similar features. Here we have seen a difference between grid and cloud layer architecture. And also see the differences of applications and services between those three computing schemes.

**Layered Architecture**

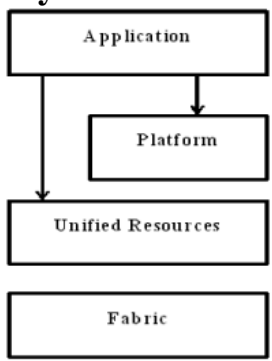

Figure: 4 Cloud computing architecture

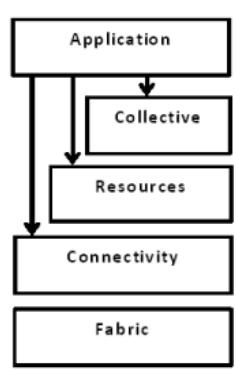

Figure: 5 Grid computing architecture

Grid uses five-layer architecture but cloud uses four-layer architecture. In both architectures first and last layers are same but collective, resources and connectivity these three layers are combined in two layers in cloud that are platform, unified resources. In grid architecture **Fabric layer** is responsible to provide resources to end users. This delivery process is managed by the grid protocols. The resources can be physical and logical also.

Physical resources like servers, catalog, storage systems etc. and logical resources like computer clusters or distributed computer pool, database system to store data, distributed file systems etc. General purpose components of fabric are GARA (General Architecture for Advanced Reservation). And specialized resource management services are Falkon. It also provides a service beyond the fabric layer [14]. **Connectivity layer** is responsible for secure network transactions. It defines an authentication protocol and core communication. GSI (Grid Security Infrastructure) is used for every grid transactions [14]. This GSI is based on a PKI which provides a single sign on authentication, communication protection, and some support for restricted delegation [15]. **Resource layer** is responsible to implement the protocols that allow the users to interact with remote resources and services. This layer builds on connectivity layer communication and authentication protocols to define protocols for secure negotiation, initiation, monitoring, control accounting, and payment of sharing operations on individual resources [15]. Two classes for the resources layer protocols are information protocols and management protocols.**Collective layer** is responsible for the simulations of multiple resources. It also contains a protocols and services and captures interactions across collection of resources [15]. Ex. MDS (monitoring and discovery services). **Application layer** is the last layer of grid architecture. It comprises the user application which constructed in a term of services and some protocols and APIs provide access to services like data access, resource discovery etc. It operates within a VO environment.

The four layer architecture of cloud computing is composed of 1.Fabric layer, 2.Unified resources, 3. Platform, 4. Application layer.

The fabric and application layer do a same functionality as grid architecture. Another difference in cluster, grid and cloud computing is showing in a table [4].

|  | Cluster | Grids | Clouds |
|---|---|---|---|
| Nodes | Tightly-coupled | Loosely-coupled | Loosely-coupled |
| Business model | No | No | Yes |
| Allocation | Centralized | Decentralized | Both |
| SLA | Limited | Yes | Yes |
| Virtualization | Little | Little | Yes |
| Heterogeneity | No | Yes | Yes |
| Areas | Educational resources, Medical search etc. | Predictive modelling and simulation, energy resources exploration etc. | Banking, Insurance, Weather forecasting, Space exploration etc. |
| Services | WMD simulation, 3D modelling, DNA sequence analysis, Molecular nanotechnology | Intergrids, Intragrids, Extragrids | IaaS, PaaS, SaaS |
| Applications | Weather modelling, Life sciences, Protein explorer, Computational fluid dynamics, Nuclear simulations, Image processing, Data mining etc. and as Internet applications are Database servers, Data mining, Email, Proxy, Security etc. | NEESgrid Earthquake Engineering Clooaboratory, The virtual Observatory World-Wide Telescope etc. | Google Docs, Basecamp, Campfire, Xero, Springpad etc. |
| Projects | Beowulf, Bereley NOW, HPVM, Solaris MC, etc. | Tera grid, Globus project, NKN, Fusion Collaboratory, EGEE etc. | CERN, Unified cloud interface(UCI), OpenNebula, CESWP, TClouds |

## VI. Conclusion

In this paper we studied features of Grid, Cluster and Cloud Computing along with the existing implementations of the same. Since these technologies have huge potential, their possible future uses are also quoted and analysed. We also found that experts in these areas will have to look for the scopes of improvements in not only the reliability, efficiency and performance, but also in fault tolerance. Modern day faults are less in frequency but very difficult to deal with. Some of the possible faults are also discussed in separate sections which are yet to be addressed by researchers.

To conclude, we can say that cluster computing is pillar architecture of cloud and grid. Many names in the market already have cloud like YouTube, Google docs etc. In coming years private computing will be in the hands of clouds. Cloud computing uses VPNs to provide services to the consumers.

Some companies which provide these services and tools and required protocols are Amazon, Microsoft, Yahoo, IBM, Google etc. But in all three of the computing technologies, some issues are like privacy, data safety and vendor lock-in. But many researchers believe that the only disadvantage of cloud computing is less security.